\input epsf

\def\INSERTCAP#1#2{\vbox{%
{\narrower\noindent%
\multiply\baselineskip by 3%
\divide\baselineskip by 4%
{\rm Table #1 }{\sl #2 \medskip}}
}}
%
%
\input hyperbasics 
%
%
\def\unredoffs{} \def\redoffs{\voffset=-.31truein\hoffset=-.48truein}
\def\speclscape{}
%
%
%
%
%
\newbox\leftpage \newdimen\fullhsize \newdimen\hstitle \newdimen\hsbody
\tolerance=1000\hfuzz=2pt
\catcode`\@=11 
\ifx\hyperdef\UNd@FiNeD\def\hyperdef#1#2#3#4{#4}\def\hyperref#1#2#3#4{#4}\fi
\def\bigans{b }
\def\answ{b }
%
\ifx\answ\bigans\message{(This will come out unreduced.}
\magnification=1200\unredoffs\baselineskip=16pt plus 2pt minus 1pt
\hsbody=\hsize \hstitle=\hsize 
\else\message{(This will be reduced.} \let\l@r=L
\magnification=1000\baselineskip=16pt plus 2pt minus 1pt \vsize=7truein
\redoffs \hstitle=8truein\hsbody=4.75truein\fullhsize=10truein\hsize=\hsbody
\output={\ifnum\pageno=0 
  \shipout\vbox{\speclscape{\hsize\fullhsize\makeheadline}
    \hbox to \fullhsize{\hfill\pagebody\hfill}}\advancepageno
  \else
  \almostshipout{\leftline{\vbox{\pagebody\makefootline}}}\advancepageno
  \fi}
\def\almostshipout#1{\if L\l@r \count1=1 \message{[\the\count0.\the\count1]}
      \global\setbox\leftpage=#1 \global\let\l@r=R
 \else \count1=2
  \shipout\vbox{\speclscape{\hsize\fullhsize\makeheadline}
      \hbox to\fullhsize{\box\leftpage\hfil#1}}  \global\let\l@r=L\fi}
\fi
%
\newcount\yearltd\yearltd=\year\advance\yearltd by -1900

\def\Title#1#2{\nopagenumbers\abstractfont\hsize=\hstitle\rightline{#1}%
\vskip 1in\centerline{\titlefont #2}\abstractfont\vskip .5in\pageno=0}
\def\Date#1{\vfill\leftline{#1}\tenpoint\supereject\global\hsize=\hsbody%
\footline={\hss\tenrm\hyperdef\hypernoname{page}\folio\folio\hss}}%
%

\def\draftmode{\message{ DRAFTMODE }\def\draftdate{{\rm preliminary draft:
\number\month/\number\day/\number\yearltd\ \ \hourmin}}%
\headline={\hfil\draftdate}\writelabels\baselineskip=20pt plus 2pt minus 2pt
 {\count255=\time\divide\count255 by 60 \xdef\hourmin{\number\count255}
  \multiply\count255 by-60\advance\count255 by\time
  \xdef\hourmin{\hourmin:\ifnum\count255<10 0\fi\the\count255}}}
\def\nolabels{\def\wrlabeL##1{}\def\eqlabeL##1{}\def\reflabeL##1{}}
\def\writelabels{\def\wrlabeL##1{\leavevmode\vadjust{\rlap{\smash%
{\line{{\escapechar=` \hfill\rlap{\sevenrm\hskip.03in\string##1}}}}}}}%
\def\eqlabeL##1{{\escapechar-1\rlap{\sevenrm\hskip.05in\string##1}}}%
\def\reflabeL##1{\noexpand\llap{\noexpand\sevenrm\string\string\string##1}}}
\nolabels
%
\global\newcount\secno \global\secno=0
\global\newcount\meqno \global\meqno=1
\def\newsec#1{\global\advance\secno by1\message{(\the\secno. #1)}
\global\subsecno=0\eqnres@t\noindent
{\bf\hyperdef\hypernoname{section}{\the\secno}{\the\secno.} #1}%
\writetoca{{\string\hyperref{}{section}{\the\secno}{\secsym}} {#1}}%
\par\nobreak\medskip\nobreak}
\def\eqnres@t{\xdef\secsym{\the\secno.}\global\meqno=1\bigbreak\bigskip}
\def\sequentialequations{\def\eqnres@t{\bigbreak}}\xdef\secsym{}
\global\newcount\subsecno \global\subsecno=0
\def\subsec#1{\global\advance\subsecno by1\message{(\secsym\the\subsecno. #1)}
\ifnum\lastpenalty>9000\else\bigbreak\fi
\noindent{\it\hyperdef\hypernoname{subsection}{\secsym\the\subsecno}%
{\secsym\the\subsecno.} #1}\writetoca{\string\quad
{\string\hyperref{}{subsection}{\secsym\the\subsecno}{\secsym\the\subsecno.}}
{#1}}\par\nobreak\medskip\nobreak}
\def\appendix#1#2{\global\meqno=1\global\subsecno=0\xdef\secsym{\hbox{#1.}}%
\bigbreak\bigskip\noindent{\bf Appendix \hyperdef\hypernoname{appendix}{#1}%
{#1.} #2}\message{(#1. #2)}\xdef\appsym{#1}%
\writetoca{\string\hyperref{}{appendix}{#1}{Appendix {#1.}} {#2}}%
\par\nobreak\medskip\nobreak}
%
%
\def\checkm@de#1#2{\ifmmode{\def\f@rst##1{##1}\hyperdef\hypernoname{equation}%
{#1}{#2}}\else\hyperref{}{equation}{#1}{#2}\fi}
\def\eqnn#1{\DefWarn#1\xdef #1{(\noexpand\relax\noexpand\checkm@de%
{\secsym\the\meqno}{\secsym\the\meqno})}%
\wrlabeL#1\writedef{#1\leftbracket#1}\global\advance\meqno by1}
\def\f@rst#1{\c@t#1a\em@ark}\def\c@t#1#2\em@ark{#1}
\def\eqna#1{\DefWarn#1\wrlabeL{#1$\{\}$}%
\xdef #1##1{(\noexpand\relax\noexpand\checkm@de%
{\secsym\the\meqno\noexpand\f@rst{##1}}{\hbox{$\secsym\the\meqno##1$}})}
\writedef{#1\numbersign1\leftbracket#1{\numbersign1}}\global\advance\meqno by1}
\def\eqn#1#2{\DefWarn#1%
\xdef #1{(\noexpand\hyperref{}{equation}{\secsym\the\meqno}%
{\secsym\the\meqno})}$$#2\eqno(\hyperdef\hypernoname{equation}%
{\secsym\the\meqno}{\secsym\the\meqno})\eqlabeL#1$$%
\writedef{#1\leftbracket#1}\global\advance\meqno by1}
\def\xeqn{\expandafter\xe@n}\def\xe@n(#1){#1}
\def\xeqna#1{\expandafter\xe@n#1}
\def\eqns#1{(\e@ns #1{\hbox{}})}
\def\e@ns#1{\ifx\UNd@FiNeD#1\message{eqnlabel \string#1 is undefined.}%
\xdef#1{(?.?)}\fi{\let\hyperref=\relax\xdef\next{#1}}%
\ifx\next\em@rk\def\next{}\else%
\ifx\next#1\xeqn#1\else\def\n@xt{#1}\ifx\n@xt\next#1\else\xeqna#1\fi
\fi\let\next=\e@ns\fi\next}

\def\DefWarn#1{\ifx\UNd@FiNeD#1\else
\immediate\write16{*** WARNING: the label \string#1 is already defined ***}\fi}
%
\newskip\footskip\footskip14pt plus 1pt minus 1pt 
\def\footnotefont{\ninepoint}\def\f@t#1{\footnotefont #1\@foot}
\def\f@@t{\baselineskip\footskip\bgroup\footnotefont\aftergroup\@foot\let\next}
\setbox\strutbox=\hbox{\vrule height9.5pt depth4.5pt width0pt}
\global\newcount\ftno \global\ftno=0
\def\foot{\global\advance\ftno by1\def\foot@rg{\hyperref{}{footnote}%
{\the\ftno}{\the\ftno}\xdef\foot@rg{\noexpand\hyperdef\noexpand\hypernoname%
{footnote}{\the\ftno}{\the\ftno}}}\footnote{$^{\foot@rg}$}}
%
\newwrite\ftfile
\def\footend{\def\foot{\global\advance\ftno by1\chardef\wfile=\ftfile
\hyperref{}{footnote}{\the\ftno}{$^{\the\ftno}$}%
\ifnum\ftno=1\immediate\openout\ftfile=\jobname.fts\fi%
\immediate\write\ftfile{\noexpand\smallskip%
\noexpand\item{\noexpand\hyperdef\noexpand\hypernoname{footnote}
{\the\ftno}{f\the\ftno}:\ }\pctsign}\findarg}%
\def\footatend{\vfill\eject\immediate\closeout\ftfile{\parindent=20pt
\centerline{\bf Footnotes}\nobreak\bigskip\input \jobname.fts }}}
\def\footatend{}
%
%
\global\newcount\refno \global\refno=1
\newwrite\rfile
\def\ref{[\hyperref{}{reference}{\the\refno}{\the\refno}]\nref}
\def\nref#1{\DefWarn#1%
\xdef#1{[\noexpand\hyperref{}{reference}{\the\refno}{\the\refno}]}%
\writedef{#1\leftbracket#1}%
\ifnum\refno=1\immediate\openout\rfile=\jobname.refs\fi
\chardef\wfile=\rfile\immediate\write\rfile{\noexpand\item{[\noexpand\hyperdef%
\noexpand\hypernoname{reference}{\the\refno}{\the\refno}]\ }%
\reflabeL{#1\hskip.31in}\pctsign}\global\advance\refno by1\findarg}
\def\findarg#1#{\begingroup\obeylines\newlinechar=`\^^M\pass@rg}
{\obeylines\gdef\pass@rg#1{\writ@line\relax #1^^M\hbox{}^^M}%
\gdef\writ@line#1^^M{\expandafter\toks0\expandafter{\striprel@x #1}%
\edef\next{\the\toks0}\ifx\next\em@rk\let\next=\endgroup\else\ifx\next\empty%
\else\immediate\write\wfile{\the\toks0}\fi\let\next=\writ@line\fi\next\relax}}
\def\striprel@x#1{} \def\em@rk{\hbox{}}
\def\lref{\begingroup\obeylines\lr@f}
\def\lr@f#1#2{\DefWarn#1\gdef#1{\let#1=\UNd@FiNeD\ref#1{#2}}\endgroup\unskip}
\def\semi{;\hfil\break}
\def\addref#1{\immediate\write\rfile{\noexpand\item{}#1}} 
\def\listrefs{\footatend\vfill\supereject\immediate\closeout\rfile\writestoppt
\baselineskip=\footskip\centerline{{\bf References}}\bigskip{\parindent=20pt%
\frenchspacing\escapechar=` \input \jobname.refs\vfill\eject}\nonfrenchspacing}
\def\startrefs#1{\immediate\openout\rfile=\jobname.refs\refno=#1}
\def\xref{\expandafter\xr@f}\def\xr@f[#1]{#1}
\def\refs#1{\count255=1[\r@fs #1{\hbox{}}]}
\def\r@fs#1{\ifx\UNd@FiNeD#1\message{reflabel \string#1 is undefined.}%
\nref#1{need to supply reference \string#1.}\fi%
\vphantom{\hphantom{#1}}{\let\hyperref=\relax\xdef\next{#1}}%
\ifx\next\em@rk\def\next{}%
\else\ifx\next#1\ifodd\count255\relax\xref#1\count255=0\fi%
\else#1\count255=1\fi\let\next=\r@fs\fi\next}
%

%
\newwrite\ffile\global\newcount\figno \global\figno=1
\def\fig{fig.~\hyperref{}{figure}{\the\figno}{\the\figno}\nfig}
\def\nfig#1{\DefWarn#1%
\xdef#1{fig.~\noexpand\hyperref{}{figure}{\the\figno}{\the\figno}}%
\writedef{#1\leftbracket fig.\noexpand~\xfig#1}%
\ifnum\figno=1\immediate\openout\ffile=\jobname.figs\fi\chardef\wfile=\ffile%
{\let\hyperref=\relax
\immediate\write\ffile{\noexpand\medskip\noexpand\item{Fig.\ %
\noexpand\hyperdef\noexpand\hypernoname{figure}{\the\figno}{\the\figno}. }
\reflabeL{#1\hskip.55in}\pctsign}}\global\advance\figno by1\findarg}
\def\listfigs{\vfill\eject\immediate\closeout\ffile{\parindent40pt
\baselineskip14pt\centerline{{\bf Figure Captions}}\nobreak\medskip
\escapechar=` \input \jobname.figs\vfill\eject}}
\def\xfig{\expandafter\xf@g}\def\xf@g fig.\penalty\@M\ {}
\def\figs#1{figs.~\f@gs #1{\hbox{}}}
\def\f@gs#1{{\let\hyperref=\relax\xdef\next{#1}}\ifx\next\em@rk\def\next{}\else
\ifx\next#1\xfig #1\else#1\fi\let\next=\f@gs\fi\next}
\def\figin{\epsfcheck\figin}\def\figins{\epsfcheck\figins}
\def\epsfcheck{\ifx\epsfbox\UNd@FiNeD
\message{(NO epsf.tex, FIGURES WILL BE IGNORED)}
\gdef\figin##1{\vskip2in}\gdef\figins##1{\hskip.5in}
\else\message{(FIGURES WILL BE INCLUDED)}%
\gdef\figin##1{##1}\gdef\figins##1{##1}\fi}
\def\DefWarn#1{}
\def\figinsert{\goodbreak\midinsert}
\def\ifig#1#2#3{\DefWarn#1\xdef#1{fig.~\noexpand\hyperref{}{figure}%
{\the\figno}{\the\figno}}\writedef{#1\leftbracket fig.\noexpand~\xfig#1}%
\figinsert\figin{\centerline{#3}}\medskip\centerline{\vbox{\baselineskip12pt
\advance\hsize by -1truein\noindent\wrlabeL{#1=#1}\footnotefont%
{\bf Fig.~\hyperdef\hypernoname{figure}{\the\figno}{\the\figno}:} #2}}
\bigskip\endinsert\global\advance\figno by1}
\newwrite\lfile
{\escapechar-1\xdef\pctsign{\string\%}\xdef\leftbracket{\string\{}
\xdef\rightbracket{\string\}}\xdef\numbersign{\string\#}}
\def\writedefs{\immediate\openout\lfile=\jobname.defs \def\writedef##1{%
{\let\hyperref=\relax\let\hyperdef=\relax\let\hypernoname=\relax
 \immediate\write\lfile{\string\def\string##1\rightbracket}}}}%
\def\writestop{\def\writestoppt{\immediate\write\lfile{\string\pageno%
\the\pageno\string\startrefs\leftbracket\the\refno\rightbracket%
\string\def\string\secsym\leftbracket\secsym\rightbracket%
\string\secno\the\secno\string\meqno\the\meqno}\immediate\closeout\lfile}}
\def\writestoppt{}\def\writedef#1{}
\def\seclab#1{\DefWarn#1%
\xdef #1{\noexpand\hyperref{}{section}{\the\secno}{\the\secno}}%
\writedef{#1\leftbracket#1}\wrlabeL{#1=#1}}
\def\subseclab#1{\DefWarn#1%
\xdef #1{\noexpand\hyperref{}{subsection}{\secsym\the\subsecno}%
{\secsym\the\subsecno}}\writedef{#1\leftbracket#1}\wrlabeL{#1=#1}}
\def\applab#1{\DefWarn#1%
\xdef #1{\noexpand\hyperref{}{appendix}{\appsym}{\appsym}}%
\writedef{#1\leftbracket#1}\wrlabeL{#1=#1}}
\newwrite\tfile \def\writetoca#1{}
\def\leaderfill{\leaders\hbox to 1em{\hss.\hss}\hfill}
\def\writetoc{\immediate\openout\tfile=\jobname.toc
   \def\writetoca##1{{\edef\next{\write\tfile{\noindent ##1
   \string\leaderfill {\string\hyperref{}{page}{\noexpand\number\pageno}%
                       {\noexpand\number\pageno}} \par}}\next}}}
\newread\ch@ckfile
\def\listtoc{\immediate\closeout\tfile\immediate\openin\ch@ckfile=\jobname.toc
\ifeof\ch@ckfile\message{no file \jobname.toc, no table of contents this pass}%
\else\closein\ch@ckfile\centerline{\bf Contents}\nobreak\medskip%
{\baselineskip=12pt\footnotefont\parskip=0pt\catcode`\@=11\input\jobname.toc
\catcode`\@=12\bigbreak\bigskip}\fi}
\catcode`\@=12 
%
\edef\tfontsize{\ifx\answ\bigans scaled\magstep3\else scaled\magstep4\fi}
\font\titlerm=cmr10 \tfontsize \font\titlerms=cmr7 \tfontsize
\font\titlermss=cmr5 \tfontsize \font\titlei=cmmi10 \tfontsize
\font\titleis=cmmi7 \tfontsize \font\titleiss=cmmi5 \tfontsize
\font\titlesy=cmsy10 \tfontsize \font\titlesys=cmsy7 \tfontsize
\font\titlesyss=cmsy5 \tfontsize \font\titleit=cmti10 \tfontsize
\skewchar\titlei='177 \skewchar\titleis='177 \skewchar\titleiss='177
\skewchar\titlesy='60 \skewchar\titlesys='60 \skewchar\titlesyss='60
\def\titlefont{\def\rm{\fam0\titlerm}
\textfont0=\titlerm \scriptfont0=\titlerms \scriptscriptfont0=\titlermss
\textfont1=\titlei \scriptfont1=\titleis \scriptscriptfont1=\titleiss
\textfont2=\titlesy \scriptfont2=\titlesys \scriptscriptfont2=\titlesyss
\textfont\itfam=\titleit \def\it{\fam\itfam\titleit}\rm}
 \ifx\answ\bigans\else scaled\magstep1\fi
\ifx\answ\bigans\def\abstractfont{\tenpoint}\else
\font\absit=cmti10 scaled \magstep1
\font\abssl=cmsl10 scaled \magstep1
\font\absrm=cmr10 scaled\magstep1 \font\absrms=cmr7 scaled\magstep1
\font\absrmss=cmr5 scaled\magstep1 \font\absi=cmmi10 scaled\magstep1
\font\absis=cmmi7 scaled\magstep1 \font\absiss=cmmi5 scaled\magstep1
\font\abssy=cmsy10 scaled\magstep1 \font\abssys=cmsy7 scaled\magstep1
\font\abssyss=cmsy5 scaled\magstep1 \font\absbf=cmbx10 scaled\magstep1
\skewchar\absi='177 \skewchar\absis='177 \skewchar\absiss='177
\skewchar\abssy='60 \skewchar\abssys='60 \skewchar\abssyss='60
\def\abstractfont{\def\rm{\fam0\absrm}
\textfont0=\absrm \scriptfont0=\absrms \scriptscriptfont0=\absrmss
\textfont1=\absi \scriptfont1=\absis \scriptscriptfont1=\absiss
\textfont2=\abssy \scriptfont2=\abssys \scriptscriptfont2=\abssyss
\textfont\itfam=\absit \def\it{\fam\itfam\absit}\def\footnotefont{\tenpoint}%
\textfont\slfam=\abssl \def\sl{\fam\slfam\abssl}%
\textfont\bffam=\absbf \def\bf{\fam\bffam\absbf}\rm}\fi
\def\tenpoint{\def\rm{\fam0\tenrm}
\textfont0=\tenrm \scriptfont0=\sevenrm \scriptscriptfont0=\fiverm
\textfont1=\teni  \scriptfont1=\seveni  \scriptscriptfont1=\fivei
\textfont2=\tensy \scriptfont2=\sevensy \scriptscriptfont2=\fivesy
\textfont\itfam=\tenit \def\it{\fam\itfam\tenit}\def\footnotefont{\ninepoint}%
\textfont\bffam=\tenbf \def\bf{\fam\bffam\tenbf}\def\sl{\fam\slfam\tensl}\rm}
\font\ninerm=cmr9 \font\sixrm=cmr6 \font\ninei=cmmi9 \font\sixi=cmmi6
\font\ninesy=cmsy9 \font\sixsy=cmsy6 \font\ninebf=cmbx9
\font\nineit=cmti9 \font\ninesl=cmsl9 \skewchar\ninei='177
\skewchar\sixi='177 \skewchar\ninesy='60 \skewchar\sixsy='60
\def\ninepoint{\def\rm{\fam0\ninerm}
\textfont0=\ninerm \scriptfont0=\sixrm \scriptscriptfont0=\fiverm
\textfont1=\ninei \scriptfont1=\sixi \scriptscriptfont1=\fivei
\textfont2=\ninesy \scriptfont2=\sixsy \scriptscriptfont2=\fivesy
\textfont\itfam=\ninei \def\it{\fam\itfam\nineit}\def\sl{\fam\slfam\ninesl}%
\textfont\bffam=\ninebf \def\bf{\fam\bffam\ninebf}\rm}
%
%

\hyphenation{anom-aly anom-alies coun-ter-term coun-ter-terms}
\def\inv{^{\raise.15ex\hbox{${\scriptscriptstyle -}$}\kern-.05em 1}}

\def\Dsl{\,\raise.15ex\hbox{/}\mkern-13.5mu D} 
\def\dsl{\raise.15ex\hbox{/}\kern-.57em\partial}

\def\lspace{\ifx\answ\bigans{}\else\qquad\fi}
\def\lbspace{\ifx\answ\bigans{}\else\hskip-.2in\fi} 
\def\boxeqn#1{\vcenter{\vbox{\hrule\hbox{\vrule\kern3pt\vbox{\kern3pt
	\hbox{${\displaystyle #1}$}\kern3pt}\kern3pt\vrule}\hrule}}}
\def\mbox#1#2{\vcenter{\hrule \hbox{\vrule height#2in
		\kern#1in \vrule} \hrule}}  
%
 \def\CO{{\cal O}} 

\def\vev#1{\langle #1 \rangle}

\def\darr#1{\raise1.5ex\hbox{$\leftrightarrow$}\mkern-16.5mu #1}

\def\roughly#1{\raise.3ex\hbox{$#1$\kern-.75em\lower1ex\hbox{$\sim$}}}
\def\OMIT#1{}

\def\spur{\raise.15ex\hbox{/}\kern-.57em }

\def\CO{{\cal O}}
\def\ccdot{\hbox{\kern-.1em$\cdot$\kern-.1em}}
\def\frac#1#2{{#1\over#2}}

\def\larr#1{\raise1.5ex\hbox{$\leftarrow$}\mkern-16.5mu #1}

%
%

%
%

\def\lsl{\spur {\kern0.1em l}}

%

%

\def\np#1#2#3{\NP{\bf #1} (#2) #3}
\def\pl#1#2#3{\PL{\bf #1} (#2) #3}

\def\physrev#1#2#3{\PR{\bf #1} (#2) #3}

\def\NP{{\it Nucl.\ Phys.\ }}
\def\PL{{\it Phys.\ Lett.\ }}
\def\PR{{\it Phys.\ Rev.\ }}

%
%
%
\catcode`\@=11 
\global\newcount\exerno \global\exerno=0
\def\exercise#1{\begingroup\global\advance\exerno by1%
\medbreak
\baselineskip=10pt plus 1pt minus 1pt
\parskip=0pt
\hrule\smallskip\noindent{\sl Exercise \the\secno.\the\exerno \/}
{\ninepoint #1}\smallskip\hrule\medbreak\endgroup}
%
%
\def\newsec#1{\global\advance\secno by1\message{(\the\secno. #1)}
\global\exerno=0%
\global\subsecno=0\eqnres@t\noindent{\bf\the\secno. #1}
\writetoca{{\secsym} {#1}}\par\nobreak\medskip\nobreak}
\def\appendix#1#2{\global\exerno=0%
\global\meqno=1\global\subsecno=0\xdef\secsym{\hbox{#1.}}
\bigbreak\bigskip\noindent{\bf Appendix #1. #2}\message{(#1. #2)}
\writetoca{Appendix {#1.} {#2}}\par\nobreak\medskip\nobreak}
\catcode`\@=12 
%
%
%
\def\INSERTFIG#1#2#3{\vbox{\vbox{\hfil\epsfbox{#1}\hfill}%
{\narrower\noindent%
\multiply\baselineskip by 3%
\divide\baselineskip by 4%
{\ninerm Figure #2 }{\ninesl #3 \medskip}}
}}%

%
\relax

\def\hqs{[1]}
\def\cleo{[2]}
\def\arg{[3]}
\def\aleph{[4]}
\def\hist{[5]}
\def\ffsdefd{(2.1)}
\def\drtone{[6]}
\def\spoiler{[7]}
\def\drttwo{[8]}
\def\zdef{(2.2)}
\def\disp{(2.3)}
\def\phidef{(2.4)}
\def\rein{[9]}
\def\morealpha{[10]}
\def\ipdefd{(2.5)}
\def\fis{(2.6)}
\def\inn{(2.7)}
\def\pp{(2.8)}
\def\tt{(2.9)}
\def\psidef{(2.10)}
\def\Idef{(2.11)}
\def\unitcirclebasis{(2.12)}
\def\basisfn{(2.13)}
\def\asum{(2.14)}
\def\match{[11]}
\def\luke{[12]}
\def\blife{[13]}
\def\vain{[14]}
\def\constraints{[15]}
\def\nrqm{[16]}
\def\man{[17]}
\def\shifman{[18]}
\def\falkneubert{[19]}
\def\uncer{(6.1)}
\def\avgv{(6.2)}


%
\def\w{{\omega}}

\def\t{{\theta}}
\Title{\vbox{\hbox{UCSD/PTH 95-03}\hbox{hep-ph/9504235}}}{\vbox{%
\centerline{Model-Independent Extraction of $|V_{cb}|$} 
\centerline{Using Dispersion Relations} }}
\centerline{C. Glenn Boyd\footnote{$^{\ast}$}{gboyd@ucsd.edu},
Benjam\'\i n Grinstein\footnote{$^{\dagger}$}{bgrinstein@ucsd.edu} and
Richard F. Lebed\footnote{$^{\ddagger}$}{rlebed@ucsd.edu}}
\bigskip\centerline{Department of Physics}
\centerline{University Of California, San Diego}
\centerline{La Jolla, California 92093-0319}
\vskip .3in
     	We present a method for parametrizing heavy meson semileptonic
form factors using dispersion relations, and from it produce a
two-parameter description of the $B \to B$ elastic form factor.  We
use heavy quark symmetry to relate this function to $\bar B\to D^* l
\bar \nu$ form factors, and extract $|V_{cb}|=0.037^{+0.003}_{-0.002}$
from experimental data with a least squares fit.  Our method
eliminates model-dependent uncertainties inherent in choosing a
parametrization for the extrapolation of the differential decay rate
to threshold.  The method also allows a description of $\bar B \to D l
\bar \nu$ form factors accurate to 1\% in terms of two parameters.
%
\Date{April 1995} 
\newsec{ Introduction}

  	A nonperturbative, model-independent description of QCD form
factors is a desirable ingredient for the extraction of
Cabibbo-Kobayashi-Maskawa parameters from exclusive meson decays.
Progress towards this goal has been realized by the development of
heavy quark symmetry\ref\hqs{N. Isgur and M. B. Wise, Phys.\ Lett.\ B
{\bf 232}, 113 (1989) and B {\bf 237}, 527 (1990) \semi E.  Eichten
and B. Hill, Phys.\ Lett.\ B {\bf 234}, 511 (1990) \semi M. B.
Voloshin and M. A.  Shifman, Yad.\ Fiz.\ {\bf 47}, 801 (1988) [Sov.\
J.\ Nucl.\ Phys.\ {\bf 47}, 511 (1988)].}, which relates and
normalizes the $\bar B \to D^* l \bar{\nu}$ and $\bar B \to D l
\bar{\nu}$ form factors in the context of the ${1\over M}$ expansion,
where $M$ is the heavy quark mass.  This normalization has been
used\nref\cleo{B. Barish {\it et al.} (CLEO Collaboration), Phys.\
Rev.\ D {\bf 51}, 1014 (1995).}\nref\arg{H. Albrecht {\it et al.}
(ARGUS Collaboration), Z.  Phys.\ C {\bf 57}, 533
(1993).}\nref\aleph{I. Scott (ALEPH Collaboration), to appear in:
Proceedings of the 27th International Conference on High Energy
Physics, Glasgow, Scotland, July 1994.}\refs{\cleo{--}\aleph}\ to
extract the value of the CKM parameter $|V_{cb}|$ by extrapolating the
measured form factor to zero recoil, where the normalization is
predicted.

  	This form factor extrapolation, necessary because the rate
vanishes at zero recoil, introduces an uncertainty in the value of
$|V_{cb}|$ due to the choice of parametrization.  Estimates of this
uncertainty obtained by varying parametrizations suffer the same
ambiguity. This ambiguity could be eliminated if one had a
nonperturbative, model-independent characterization of the form factor
in terms of a small number of parameters.

  	In this paper we use dispersion relations to derive such a
characterization and apply it towards the extraction of $|V_{cb}|$.
The characterization uses two parameters that describe the form factor
over the entire physical range to 1\% accuracy. For computational
convenience we use heavy quark symmetry in our characterization, but
other than the normalization at threshold, this is an inessential
ingredient which may be discarded at the cost of some extra algebra.

  	In Sec.\ 2 we describe a well-known method\ref\hist{N. N.
Meiman, Sov.\ Phys.\ JETP {\bf 17}, 830 (1963) \semi S. Okubo and I.
Fushih, Phys.\ Rev.\ D {\bf 4}, 2020 (1971) \semi V. Singh and A. K.
Raina, Fortschritte der Physik {\bf 27}, 561 (1979) \semi C. Bourrely,
B. Machet, and E. de Rafael, Nucl.\ Phys.\ {\bf B189}, 157 (1981).}
for using QCD dispersion relations and analyticity to place
constraints on hadronic form factors.  We then derive a basis for
functions that obey the constraints imposed on the $B \rightarrow B$
elastic form factor $F$, and show that to 1\% accuracy, only two terms
in the basis function expansion need be kept.  In Sec.\ 3 we use heavy
quark symmetry to relate $F$ to the Isgur-Wise function, which
describes the form factors for $\bar B \to D^* l \bar{\nu}$ in the
infinite quark mass limit. We make a least squares fit to CLEO\cleo,
ARGUS\arg, and ALEPH\aleph\ data using $|V_{cb}|$ and our two basis
function parameters as variables, and present our results.
Reliability of the method is discussed in Sec.\ 4, implications for
$|V_{ub}|$ are discussed in Sec.\ 5, and concluding remarks are
presented in the final section.
\newsec{The Analyticity Constraints}
      
      Consider the form factor $F$, defined by
\eqn\ffsdefd{ 
\vev{B(p')| V_\mu | B(p)} = F(q^2)(p+p')_\mu ,
}
\nref\drtone{E.  de
Rafael and J. Taron, Phys.\ Lett.\ B {\bf 282}, 215 (1992).}%
\nref\spoiler{ E. Carlson, J. Milana, N. Isgur, T.
Mannel, and W. Roberts, Phys.\ Lett.\ B {\bf 299}, 133 (1993)\semi A.
Falk, M. Luke, and M. Wise, Phys.\ Lett.\ B {\bf 299}, 123 (1993)
\semi B. Grinstein and P. Mende, Phys.\ Lett.\ B {\bf 299}, 127 (1993)
\semi J. K\"orner and D. Pirjol, Phys.\ Lett.\ B {\bf 301}, 257
(1993).}%
\nref\drttwo{E. de Rafael and J. Taron, Phys.\ Rev.\ D {\bf
50}, 373 (1994).}%
where $V_\mu=\bar b \gamma_\mu b$, and $q^2 = (p-p')^2$ is the
momentum transfer squared. A dispersion relation for the two point
function $\vev{0|TV_\mu V_\nu|0}$ connects its perturbative evaluation
with a sum over positive-definite terms. This sum includes a
contribution $\sim|\vev{0|V_\mu|B\bar B}|^2$ which, by crossing, is
given by the analytic continuation of $F$. This procedure leads to a
bound\refs{\drtone{--}\drttwo} on a weighted integral of $|F|^2$ over
$q^2 > 4 M^2$.

	A key ingredient in this approach is the transformation that
maps the complex $q^2$ plane onto the unit disc $|z| \leq 1$:
\eqn\zdef{
\sqrt{1-{q^2 \over 4M_B^2}} = {{1+z}\over{1-z}}~.
}
In terms of the angular variable $e^{i \theta} \equiv z$, the
once-subtracted QCD dispersion relation may be written as
\eqn\disp{
{1\over2\pi} \int_0^{2\pi} d\theta\, |\phi(e^{i \theta}) F
(e^{i \theta})|^2 \leq {1\over\pi}~,
}
where the weighing function $\phi(z)$ contains both the Jacobian of
the variable transformation and the essential physics of the
perturbative QCD calculation\drtone:
\eqn\phidef{
\phi (z) = {1 \over 16}\sqrt{{5 n_f} \over 6 \rho} (1 +z)^2 \sqrt{1-z} .
}
\nref\rein{L. J. Reinders, H. R. Rubinstein, and S. Yazaki,
Nucl.\ Phys.\ {\bf B186}, 109 (1981).}%
\nref\morealpha{M. A. Shifman, A. I. Vainshtein, M. B. Voloshin, and
V. I.  Zakharov, Phys.\ Lett.\ B {\bf 77}, 80 (1978) \semi M. A.
Shifman, A.  I. Vainshtein, and V. I. Zakharov, Nucl.\ Phys.\ {\bf
B147}, 385 (1979).}%
Here $n_f$ is the number of light flavors for
which $SU(n_f)$ flavor symmetry is valid; we take $n_f =2$.
Perturbative corrections to the dispersion relation are incorporated
in $\rho$, which has been computed\refs{\rein{,}\morealpha} to
$\CO(\alpha_s)$, $\rho = 1 + 0.73
\alpha_s(m_b) \approx 1.20$.

	In terms of the inner product defined by
\eqn\ipdefd{
(f,g)\equiv{1\over2\pi}\int_0^{2\pi}d\theta\, f^*(\theta)g(\theta)~,
}
and the function 
\eqn\fis{
\Psi(\t) = \phi(e^{i\t}) F(e^{i\t})~,
}
the dispersion relation Eq.~\disp\ reads
\eqn\inn{
(\Psi,\Psi)\le{1\over\pi}~.
}
Physically, poles of $F$ inside the unit disc originate from
resonances below threshold and cannot be ignored\spoiler; for the $B$
system these are the resonances $\Upsilon_{1,2,3}$.  A simple but
effective trick\drttwo\ eliminates the poles with no reference to the
size of their residues but rather only their positions ({\it i.e.},
masses).  Define the function
\eqn\pp{
P(z) = {(z -z_1)(z -z_2)(z -z_3) \over
        (1 - \bar z_1 z)(1 - \bar z_2 z)(1 - \bar z_3 z)},
}
where the $z_i$ correspond to the values $q^2 = M^2_{\Upsilon_i}$,
and the constant
\eqn\tt{
T = - P(0) \phi(0) F(0) .
}
By $b$-number conservation, $ F(0) = 1$.  Since $P(z)$ has modulus one
on the unit circle, the function
\eqn\psidef{
f_0(z) = {1\over z} [\phi(z) P(z) F(z) + T] 
}
is analytic on the unit disk and obeys
\eqn\Idef{
(f_0,f_0) \le {1 \over \pi} - |T|^2 \equiv I.  
}
Any function $g(z)$ that is analytic in the unit disc and obeys
Eq.~\Idef\ may be expanded in an orthonormal basis
\eqn\unitcirclebasis{
g(z) = \sum_{n=0}^{\infty} a_n z^n  ,
}
where $a_n = (z^n,g)$.
The QCD form factor may therefore be written as
\eqn\basisfn{
F(z) = {1\over P(z) \phi(z)} \left[ 
    \sum_{n=0}^{\infty} a_n z^{n+1} - T\right],
}
with 
\eqn\asum{
\sum_{n=0}^{\infty} 
|a_n|^2 \le {1 \over \pi} - |T|^2.
}

In the next section we will use heavy quark symmetries to relate $F$
to the form factors for $\bar B\to D^* l\bar\nu$, where the physical
kinematic range is $0 < z < 0.056$. Thus, retaining only $a_0$ and
$a_1$ induces a maximum relative error of ${\sqrt{I}(0.056)^3 / P(0)
\phi(0)} \approx 0.009 $.
\newsec{Extraction of $|V_{cb}|$}

\subsec{Heavy Quark Symmetry Relations}
	In the infinite $b$ and $c$ quark mass limit all the form
factors for $\bar B \to D l \bar \nu$ and $\bar B \to D^* l \bar \nu$
are given by one universal ``Isgur-Wise'' function.  This allows us to
apply the constraint on $F$ to the particular combination of form
factors actually measured, rather than deriving constraints for each
form factor separately.  The Isgur-Wise function is related to the
form factor $F$ by $F(\w) = \eta_B \xi(\w)$, where $\w \equiv v \cdot
v'$, and the short-distance matching correction $\eta_B$ is unity at
threshold by the Ademollo-Gatto theorem.  To compare with
data, we need the analogous short-distance matching and running
correction\ref\match{A. F.  Falk, et al, \np{B343}{1990}{1}\semi
A. F.  Falk and B. Grinstein, \pl{B247}{1990}{406}\semi
A. F. Falk and B. Grinstein, \pl{B249}{1990}{314}\semi
M. Neubert, SLAC Report No.\ SLAC-PUB-6263.}
for $\bar B \to D^*$ form factors.
For example, for the vector current form factor $g$ one may write
$g={\eta_D\over\eta_B}F$. This relation generally holds to order
$1/M$, but at threshold it holds to 
order~$1/M^2$\ref\luke{M. E. Luke, 
Phys.\ Lett.\ B {\bf 252}, 447 (1990).}. We treat
$\eta_D/\eta_B$ as approximately constant and equal to $0.985$. The
errors from this approximation should be no larger than the neglected
$1/M$ corrections.

\subsec{Maximum Likelihood Fit}

        Once the essential physics of QCD is incorporated into the
calculation via Eqs.~\basisfn\ and~\asum, the maximum likelihood fit
is simply an ordinary chi-squared minimization with parameters
$|V_{cb}|$ and the basis coefficients $\{a_n\}$.  As mentioned
previously, the smallness of $z$ in the physical range allows us to
ignore all coefficients except $a_0$ and $a_1$.
We normalize input data to a $B$
lifetime\ref\blife{W. Venus, in: Lepton and Photon Interactions, XVI
International Symposium, eds. Persis Drell and David Rubin, AIP Press,
New York (1994).} of $\tau_B = 1.61$ ps.

\bigskip
\vbox{\medskip
\hfil\vbox{\offinterlineskip
\hrule
\halign{&\vrule#&\strut\quad\hfil$#$\quad\cr
height2pt&\omit&&\omit&&\omit&&\omit&\cr
&|V_{cb}|\cdot 10^3&&a_0\hfil&&a_1\hfil&&\rm{Expt.}\hfil&\cr
height3pt&\omit&&\omit&&\omit&&\omit&\cr
\noalign{\hrule}
height2pt&\omit&&\omit&&\omit&&\omit&\cr
&35.7_{-2.8}^{+4.2}&&-0.01_{-0.06}^{+0.02}&&-0.55_{-0.0}^{+0.9}&&
\rm{CLEO}\cleo &\cr &47.6_{-11.2}^{+7.9}&&-0.11_{-0.02}^{+0.10}
&&0.55_{-1.1}^{+0.0}&& \rm{ARGUS}\arg &\cr
&38.0_{-4.8}^{+5.2}&&0.03_{-0.05}^{+0.04}&&-0.55_{-0.0}^{+0.5}&&
\rm{ALEPH}\aleph &\cr }
\hrule}
\hfil}
\medskip
\INSERTCAP{1}{Fit values
for $|V_{cb}|$, $a_0$, and $a_1$ from the various experiments.}
\bigskip

     	Table 1 shows the central values and $68\%$ confidence levels
for $|V_{cb}|$, $a_0$, and $a_1$ from the various experiments. The
saturation of its QCD bound by $a_1$ is not significant because its
variance is large, which arises because the contribution of $a_1$ is
suppressed by an extra power of $z$.

\INSERTFIG{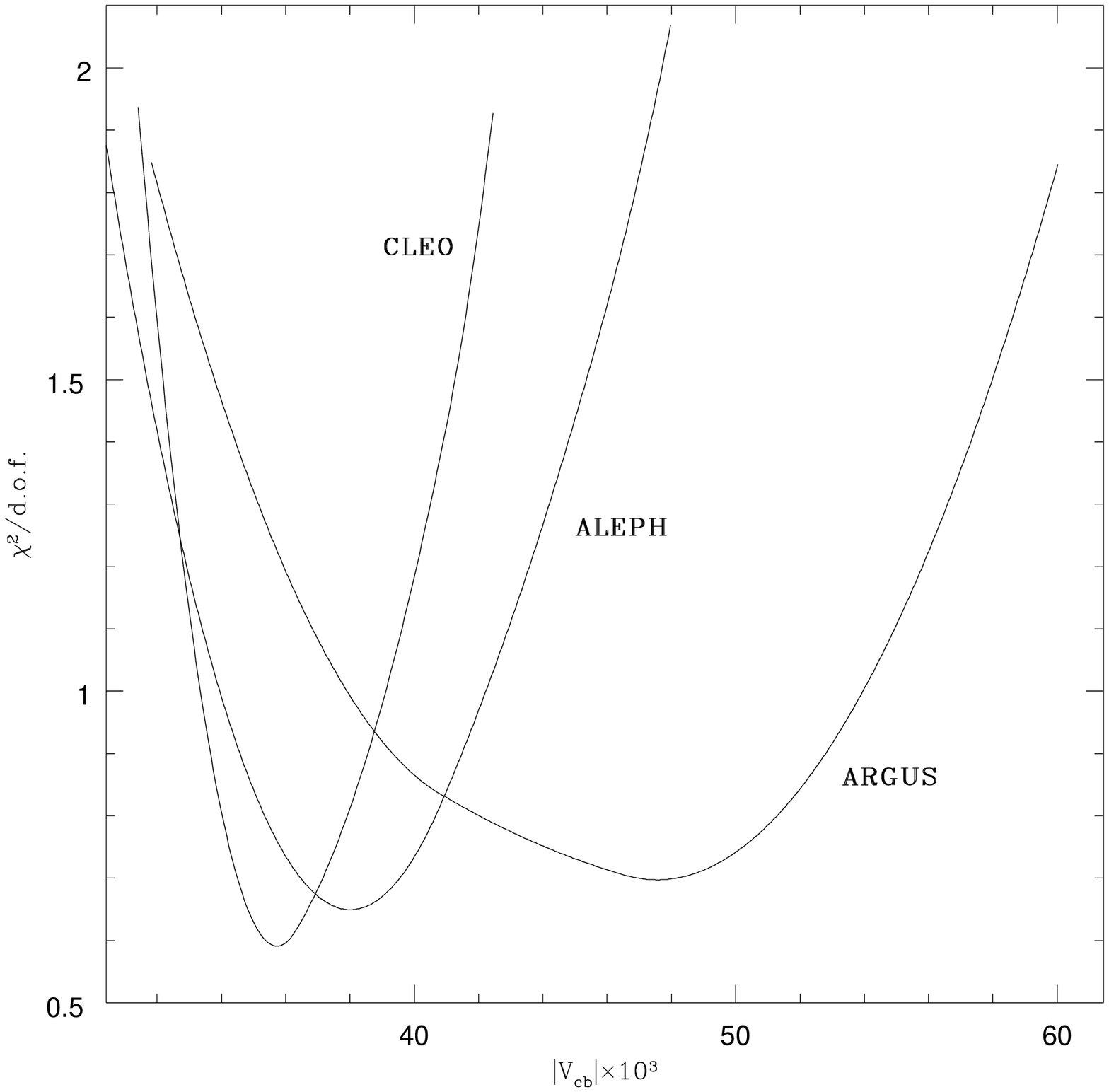}{1}{$\chi^2$ per degree of freedom resulting from
a least squares fit in $|V_{cb}|$, and $a_0$ and $a_1$ from
Eq.~\basisfn\ to CLEO, ALEPH, and ARGUS data, plotted against
$|V_{cb}|$.}

	Plots of $\chi^2$ per degree of freedom versus $|V_{cb}|$ for
each of the experiments are shown in Fig.\ 1. The minimum $\chi^2$ is
consistently low, remarkable agreement for a first-principles
parametrization.

   	Figure 2 shows the product of the best fit form factors with
$|V_{cb}|$, superimposed with experimental data.  At 90\% confidence
level, $a_0$ and $a_1$ are consistent with zero, suggesting the
dispersion relation may be saturated entirely by higher states.

	The errors on $|V_{cb}|$ in Table\ 1 are statistical only; the
treatment of systematic errors depends both on our parametrization and
a detailed understanding of the experiment.  The error implicit in the
variation over choices of parametrization, however, is absent. For the
ARGUS experiment, varying over four possible parametrizations induced
a spread of 0.012 in $|V_{cb}|$, and was the major impediment in using
heavy quark symmetry to obtain a model-independent extraction.  Even
the experiment with highest statistics, CLEO, remains sensitive to the
choice of extraction. For a linear fit, they find $|V_{cb}|\cdot 10^3
= 35.1_{-1.9}^{+1.9} $, while for a quadratic fit, they find
$|V_{cb}|\cdot 10^3= 35.3_{-3.2}^{+3.0} $. Presumably a higher-order
fit would yield even larger variances. Fortunately, the basis function
approach does not yield statistical errors indicative of such a
higher-order fit.

\INSERTFIG{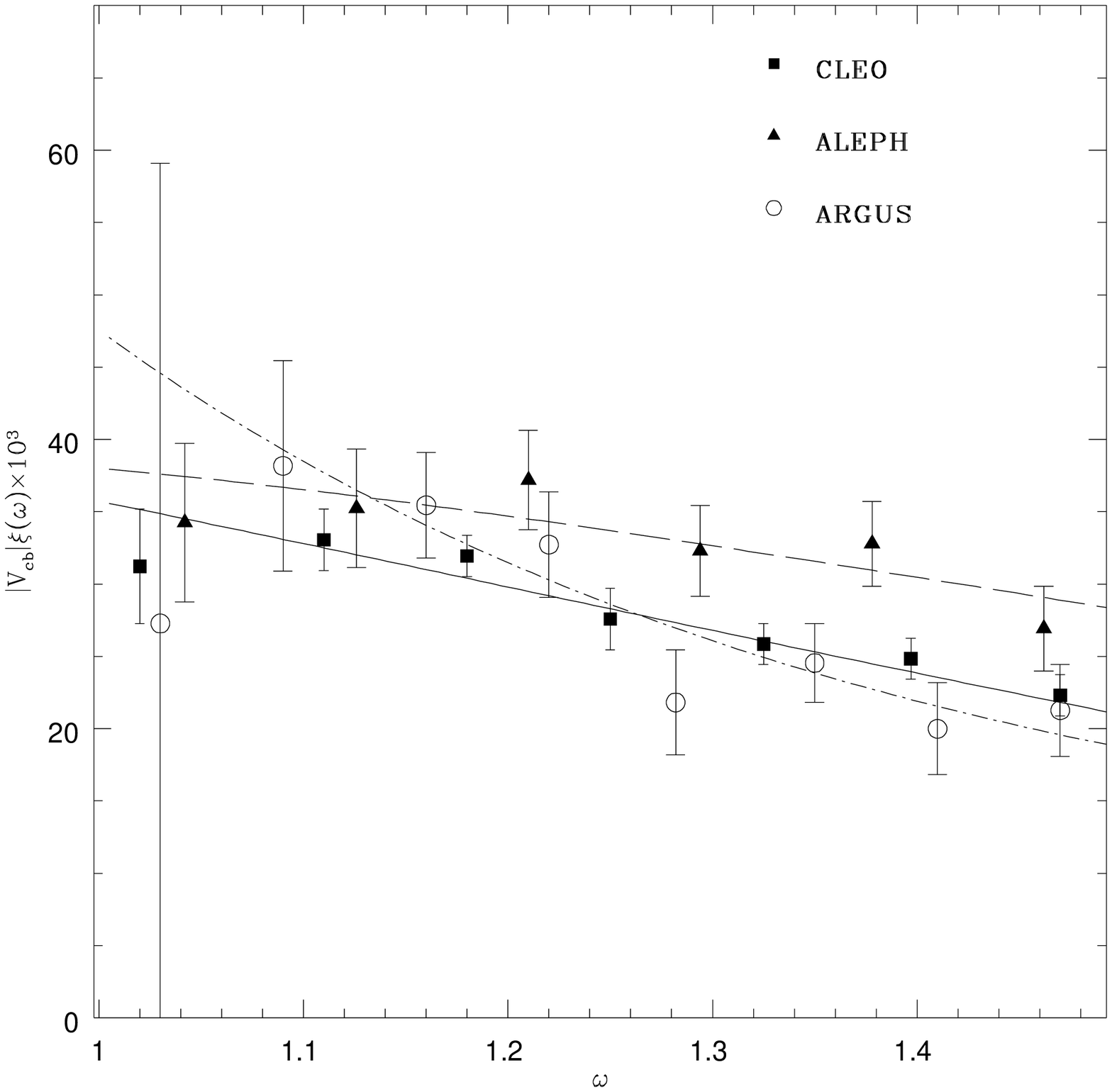}{2}{Best fit values for the product of $|V_{cb}|$
with the Isgur-Wise function for CLEO (solid line), ARGUS (dot-dashed
line), and ALEPH (dashed line) data. The data for each experiment,
adjusted for the $B$ lifetime and zero-recoil normalization used in
the text, is superimposed.}

\newsec{Discussion of the Method}
\subsec{Reliability of the Parametrization}

    	The parametrization Eq. \basisfn\ was used to make a
three-parameter ($a_0, a_1$ and $|V_{cb}|$) maximum likelihood fit to
the data.  We can list four types of correction to this equation.

    	First, perturbative $\CO(\alpha_s^2(m_b))$ corrections to the
dispersion relation \Idef\ arise as $z$-independent renormalizations
of the function $\phi$, and may be calculated systematically through
higher loop diagrams.  The extraction of $|V_{cb}|$ is rather
insensitive to such corrections: Altering by hand the perturbative
computation by $\pm 10\%$ changes the central value of $|V_{cb}|$ by
less than $\pm 0.3\%$.

\nref\vain{A. I. Vainshtein, V. I. Zakharov, and M. A. Shifman, JETP
Lett.\ {\bf 27}, 55 (1978).}

    	Second, non-perturbative corrections to the dispersion
relation may be analyzed via an operator product expansion. The first
correction takes the form of a gluon condensate, estimated to
be\refs{\morealpha{,}\vain} ${ \alpha_s(M_B) \over M_B^4 \pi}
<G_{\mu\nu}G^{\mu\nu}> \approx 10^{-5}$, which is completely
negligible.

    	Third, the truncation of Eq.~\basisfn\ at finite $n$
introduces an error proportional to ${1\over P(0) \phi(0)}
\sum_{i=n+1}^{\infty} a_{i} z^{i+1}$ and additionally suppressed 
by the coefficients $a_n$ themselves, whose sum is bounded:
$\sum_{n=0}^{\infty} |a_n|^2 \le I \approx 0.31$.  Again, for the case
at hand, the first two terms are sufficient to describe any form
factor allowed by the dispersion relations to within 1\%.

   	Finally, the application of the $B \to B$ dispersion relation
to $\bar B \to D^* l \bar \nu$ decays relies on heavy quark symmetry,
which is computationally convenient but unnecessary. The weighing
function $\phi(z)$ appropriate to $\bar B \to D l \bar \nu$ may be
readily deduced from an analogous computation for $\bar B \to \pi l
\bar \nu$\ref\constraints{C. G. Boyd, B. Grinstein, and R. F. Lebed,
U. C. San Diego Report No.\ UCSD/PTH-94/27 [hep-ph/9412324]
(unpublished).}.  Eq.~\basisfn\ then holds with the function $P(z)$
altered to reflect $B_c$ poles below threshold, the positions of which
have been calculated in the context of a nonrelativistic potential
model\ref\nrqm{ E. J. Eichten and C. Quigg,
\physrev{D49}{1994}{5845}\semi M. Baker, J. S. Ball, and F.
Zachariasen, \physrev{D45}{1992}{910}\semi R. Roncaglia {\it et al.},
\physrev{D51}{1995}{1248}\semi S. Godfrey and N. Isgur,
\physrev{D32}{1985}{189}.  }.  It should also be possible to derive
analogous dispersion relations for each of the $\bar B \to D^* l \bar
\nu$ form factors, again using no assumptions about heavy quark
symmetry.  This should allow the extraction of $|V_{cb}|$ from both
$\bar B\to D^* l \bar \nu$ and $\bar B \to D l \bar \nu$ decays, given
only the normalization of the form factors at zero recoil.

	Alternatively, one may want to use heavy quark symmetry to
relate the $\bar B\to D^* l \bar \nu$ and $\bar B \to D l \bar \nu$
form factors and construct a QCD constraint on only one of them.  Such
heavy quark symmetry relations are spoiled only by spin-symmetry
violating corrections, which are expected to be smaller than the
flavor-symmetry violating corrections inherent in the method of Sec.\
3.

	All of the errors described above are either extremely small,
or amenable to systematic reduction.  We see no theoretical obstacle
to predicting the $\bar B \to D$ and $\bar B \to D^*$ form factors to
1\% accuracy, given the normalization at threshold and sufficiently
precise measurements to fix two parameters.  Such a prediction would
not only test our understanding of QCD at an unprecedented level; it
would present a precision probe of non-standard model physics in a
hadronic arena.

\subsec{Reliability of the Extraction}

	One type of heavy quark symmetry correction arises from the
application of the $B \to B$ dispersion relation to $\bar B \to D^* l
\bar \nu$. We estimate such corrections by making a $20\%$ change in
the ranges of $a_0$ and $a_1$, resulting in a $2\%$ shift in the
central value of $|V_{cb}|$.
  
	Another heavy quark correction arises at $\CO({1\over M^2})$
in the normalization of the Isgur-Wise function at threshold.  The
normalization of the form factor $g(\w=1)$ has been estimated to be
$g(1)= 0.96$\ref\man{T. Mannel, Phys.\ Rev.\ D {\bf 50}, 428 (1994).},
$g(1)= 0.89$\ref\shifman{M. Shifman, N. Uraltsev, and A.  Vainshtein,
Univ.\ of Minn.\ Report No.\ TPI-MINN-94/13-T [hep-ph/9405207]
(unpublished).}, and $g(1)= 0.93$\ref\falkneubert{A. F.  Falk and M.
Neubert, Phys.\ Rev.\ D {\bf 47}, 2695 and 2982 (1993)
\semi M.  Neubert, Phys.\ Lett.\ B {\bf 338}, 84 (1994).}. We have
included a QCD correction of $0.985$, so to good approximation, this
simply rescales the values of $|V_{cb}|$ in Table\ 1 by ${0.985 \over
g(1)}$.

    	There are other errors in our extraction that are not purely
theoretical. The most pressing of these involve the binning of the
measured rate against $\omega$, smearing of $\omega$ introduced by
boosting from the lab to the center of mass frame, and correlation of
errors.  Randomly varying input values of $\omega$ in our least
squares fit of the CLEO data by $\pm 0.05$ changes the central value
of $|V_{cb}|$ by less than $1\%$.  A more thorough extraction can be
done by the experimental groups themselves, using our basis function
expansion in their maximum likelihood programs.

\newsec{Implications for $|V_{ub}|$}

	The parametrization of form factors in terms of our basis
functions applies to other heavy hadron decays, including $\bar B \to
\pi l \bar \nu$. In this case the range of the kinematic variable is
larger, $0< z< 0.5$, so more coefficients $a_n$ are needed for
comparable accuracy.  We expect six to eight $a_n$ will be necessary
for accuracy of a few percent over the entire kinematic range,
depending on the form of the actual data.

	A least squares fit will only produce best values of the
products $|V_{ub}| \xi(1)$ and $|V_{ub}| a_n$, destroying hopes of a
model-independent extraction of $|V_{ub}|$. However, small values of
$|V_{ub}|$ tend to wash out the nontrivial $z$ dependence, while the
$a_n{}'s$ cannot compensate because they are bounded from above, so
the extraction of a model-independent lower bound on $|V_{ub}|$ should
be possible.

	To obtain an upper bound on $|V_{ub}|$ will require as input
the overall normalization of the relevant form factor. Practically
speaking, this means using a model or lattice simulation.  Any
candidate model must predict a form factor that is consistent with the
basis functions, in the domain of validity of the model. This is a
severe test to pass\constraints, and should serve as an effective
discriminator for models.

\newsec{ Conclusions} 

    	The extraction of the CKM mixing parameter $|V_{cb}|$ involves
several types of uncertainties. Typically, these uncertainties are
classified as
\eqn\uncer{
|V_{cb}| = V \pm \{ \it{stat} \} \pm \{ \it{syst} \}
\pm \{ \it{life} \} \pm \{ \it{norm} \} \pm \{
\it{param}
\}
}
where {\it stat} and {\it syst} refer to statistical and systematic
experimental uncertainties, {\it life} refers to uncertainties in the
$B$ lifetime, {\it norm} refers to uncertainty in the value of the
form factor at threshold, and {\it param} refers to uncertainty in the
extrapolation of the measured differential rate to threshold.

   	Not only the central value, but also the statistical
uncertainty depends on the parametrization.  For example, linear fits
to CLEO data yield substantially smaller statistical uncertainties
than quadratic fits. Typically, quoted values correspond to the
parametrization yielding the smallest statistical uncertainty, in
effect throwing some statistical uncertainty into the parametrization
uncertainty, which remains implicit. Clearly, this does not improve
the accuracy with which we know $|V_{cb}|$.

  	In this paper, we have essentially eliminated the uncertainty
in the choice of parametrization.  This was accomplished in four
stages.  First, we used QCD dispersion relations to constrain the $B
\to B$ elastic form factor.  Second, we derived a set of parametrized
basis functions which automatically satisfies the dispersion
relation constraint, and expressed the $B \to B$ form factor in terms
of this basis. This expression involved an infinite number of
parameters $a_i$ bounded by $\sum_{n=0}^{\infty} |a_n|^2 \le I$.
Third, we used heavy quark symmetry to relate the $B \to B$ form
factor to $\bar B \to D^* l \bar
\nu$ form factors and fixed the normalization at threshold.  Over the
entire kinematic range relevant to $\bar B \to D^* l \bar \nu$, we
showed that neglecting all but the first two parameters $a_{0}$, $a_{1}$
resulted in at most a $1\%$ deviation in the predicted form factor.
Finally, we made a least squares fit of the differential $\bar B \to
D^* l \bar \nu$ rate to $|V_{cb}|$, $a_0$, and $a_1$.

   	The results of this fit improve on all previous extractions in
one important way: The uncertainty due to the choice of
parametrization is under control, and of order $1\%$.  Our statistical
errors are larger than many quoted values. This does not reflect an
inferiority of our method, but rather quantifies uncertainties that
were previously left implicit.  An averaged value from CLEO, ARGUS,
and ALEPH data is
\eqn\avgv{
|V_{cb}| = 0.037^{+0.003}_{-0.002}\, (\it{stat}).}
An estimation of systematic uncertainties requires a detailed
knowledge of the experiments.

   	The basis function parametrization described here allows
generalization to $\bar B \to \pi l \bar \nu$ as well. In this case, we
expect to be able to extract a lower bound on $|V_{ub}|$. In addition,
precision tests of QCD-predicted form factors are now possible; these
should be useful as checks of QCD models and lattice simulations.  As
experiments improve, they may even be used as probes of new physics.
Applications and generalizations of the methods described here look
promising.

\vskip1.2cm
{\it Acknowledgements}\hfil\break We would like to thank Hans Paar and
Persis Drell for useful discussions of the CLEO experiment.  The
research of one of us (B.G.)  is funded in part by the Alfred P. Sloan
Foundation. This work is supported in part by the Department of Energy
under contract DOE--FG03--90ER40546.

\vfill\eject
\listrefs
\bye